\documentclass[12pt]{iopart}

\usepackage{iopams}   
\usepackage{braket}
\usepackage{graphicx}
\usepackage{subfig}
\usepackage[countmax]{subfloat}
\usepackage{xcolor}

\begin{document}

\title[]{Chiral and flavor oscillations in a hyperentangled neutrino state}

\author{Victor A. S. V. Bittencourt$^{1,*}$, Massimo Blasone$^2$  and Gennaro Zanfardino$^3$}

\address{$^1$ ISIS (UMR 7006), Universit\'{e} de Strasbourg, 67000 Strasbourg, France}
\address{$^2$ Dipartimento di Fisica, Universit\`a degli Studi di Salerno, Via Giovanni Paolo II, 132 84084 Fisciano, Italy \& INFN, Sezione di Napoli, Gruppo Collegato di Salerno, Italy}
\address{$^3$ Dipartimento di Ingegneria Industriale, Universit\`a di Salerno, Via Giovanni Paolo II, 132 I-84084 Fisciano (SA), Italy \& INFN, Sezione di Napoli, Gruppo collegato di Salerno, Italy}
\address{$^*$\textbf{E-mail}: sant@unistra.fr}

\vspace{10pt}
\begin{indented}
\item[]\today
\end{indented}

\begin{abstract}
In addition to flavor oscillations, Dirac neutrinos also undergo the so-called chiral oscillations, a consequence of the free-particle dynamics under the Dirac equation. Such a transition between different chiralities affect the flavor transitions, but also can generate non-trivial correlations between the internal degrees of freedom of the particle. In this paper, we show that the state of a massive oscillating neutrino produced by weak interaction process, is an hyperentangled state, in which flavor, chirality, and spin exhibit non-trivial correlations. Using complete complementarity relations, we show that both chiral and flavor oscillations redistribute correlations and coherence in time among different partitions of the system. In a similar way, we consider a spin entangled lepton-antineutrino pair and show that there is a dynamical redistribution of spin-spin entanglement into correlations and coherence between the other degrees-of-freedom. Our analysis provides a complete characterization of the quantum correlations involved in lepton-antineutrino pairs and in single particle neutrino evolution, and provides a further insight on possible routes to interpret and measure chiral oscillations.
\end{abstract}

\section{Introduction}
\label{intro}

Quantum correlations are intrinsically present in particle physics systems: their analysis can both shine a light on fundamental phenomena, and open the way to use elementary particles for implementing quantum information protocols \cite{intro01}- \cite{meson}. One of the first examples shown in the literature are Kaon states produced in $\phi$-meson decays \cite{RMeson_BK, meson}, which exhibit entangled strangeness, that have been explored in experiments with particle accelerators.  Other particle systems in which quantum correlations have been investigated include decay and scattering processes, which can generate entangled spin states \cite{Cervera-Lierta:2017tdt}-\cite{ATLAS:2023fsd}.

Much attention has been devoted to neutrinos: as a consequence of being massive and mixed, neutrinos exhibit flavor oscillations \cite{R15,R16}, which can be described in terms of (time dependent) entanglement among different flavor modes \cite{Blasone:2007wp, Blasone:2007vw}.  It is possible to understand the flavor entanglement oscillations as a trade-off of quantum coherence shared among the flavors \cite{Blasone:2021cau,R14}, which have also been shown to be related to a violation of the Leggett-Garg inequalities \cite{R_Leggett_Gard}-\cite{LGB01}. In general, different quantum correlations play a role in flavor oscillations
\cite{QCProb}-\cite{EntropicBlas}, which are then connected with a quantum information framework to describe particle mixing \cite{Banarjee}-\cite{InfB01}. Besides flavor, Dirac neutrinos also carry two other discrete degrees-of-freedom (DoFs): spin (or helicity) and chirality \cite{R_Pal}, which are naturally encoded in the bispinorial structure of the Dirac equation. Charged weak processes that generate neutrinos involve projections into states with definite chirality, but chirality is not conserved under the free evolution of the state, which yields chiral oscillations \cite{A01}. In lepton-antineutrino states produced via weak decay, chiral oscillations redistribute quantum correlations initially encoded in the spins to the other DoFs \cite{R18,Bittencourt:2023brw}.

The interplay between flavor and chiral oscillations leads to correction terms to the flavor oscillation formula which include those obtained within  a quantum field theory treatment \cite{bla95,bla98}. Such a modification of the flavor oscillation formula is important, for example, for measurements of the cosmic neutrino background: for such non-relativistic neutrinos, chiral oscillations produce a signification depletion in the measured flux of neutrinos \cite{R17,GePasquini}. Both flavor and chiral oscillations can generate a trade-off between quantum correlations shared among the different DoFs, which in turn yields the possibility of exploring such correlations for measuring fundamental features of such particles, since chiral oscillations can behave different whether neutrinos are Dirac or Majorana particles \cite{Majorana_neu2021, Li:2023iys}. Furthermore, flavor and chiral oscillations can also influence the correlation profile of lepton-antineutrino pairs, which can potentially be exploited for measuring such fundamental phenomena \cite{R18}.

In this paper, by means of the so-called complete complementarity relations (CCRs) \cite{R_CCR_qudit}-\cite{R12}, we study the evolution in time of quantum correlations in a lepton-antineutrino pair. We describe the temporal evolution by means of the Dirac equation, such that each particle carries spin and chirality DoFs, and we include flavor mixing in the antineutrino dynamics. We show that the temporal evolution leads to a redistribution of quantum correlation among the different internal DoFs. The time-evolved state is a multipartite entangled state, referred as a {\it hyperentangled state} \cite{R20}. Then, by means of CCRs, we quantify the trade-off between quantum entanglement and quantum coherence shared between the spins of the particles, which allows us to identify in such spin-correlations, the effects of flavor and chiral oscillations. Our results provides a complete description of spin quantum correlations in lepton-antineutrino pairs.

The paper is organized as follows. In Section \ref{S2} we review two flavor neutrino oscillations using the CCRs. The neutrino is described as a composite system in an hyperentangled state formed by its intrinsic DoFs. In Section \ref{S3}, by means of the same formalism, we analyse the redistribution in time of the same CCRs quantities for the spin part of a lepton-antineutrino pair produced by a charged weak process. 

\section{CCR for two flavor neutrino oscillations}
\label{S2}

In this Section, we first briefly describe complete complementarity relations and then use them for the characterization of neutrino flavor oscillations for the simplest case of two flavors.

\subsection{Complete complementarity relations}
The so-called complete complementarity relations (CCRs) \cite{R_CCR_qubit} link different quantum correlations encoded in a quantum system, providing a useful tool for studying the interplay of correlations such as entanglement and quantum coherence. Given a pure bipartite two-qubit state, the CCR for the subsystem $k$ reads as
\begin{equation}
\label{CCR for pure qubit}
\mathcal{P}_{k}^{2}+\mathcal{V}_{k}^{2}+\mathcal{C}^{2} = 1 \qquad k=1,2
\end{equation}
where $\mathcal{P}_{k}$ is the predictability of the system which is related to the \textit{a priori} (without performing measurement) particle-like behaviour of the system. $\mathcal{V}_{k}$ is the visibility, related to the coherence of the system and $\mathcal{C}$ is the concurrence \cite{R_concurrence} which quantifies the entanglement between the subsystem $k$ and the other one.

The CCR were generalized to a pure tri-partite quantum system \cite{R_CCR_qudit}, and have been given an entropic form for multi-partite pure \cite{R11} and mixed states \cite{R12}. If the total state of the system is pure, the entropic CCR reads as
\begin{equation}
\label{CCRPURO}
\mathcal{C}_{re}(\rho^{k}) + \mathcal{P}_{vn}(\rho^{k}) + \mathcal{S}_{vn}(\rho^{k}) = \log_{2}d_{k}
\end{equation}
where the $\mathcal{S}_{vn}(\rho^{k})$ is the von Neumann entropy, it is related to the entanglement between subsystem $k$ and the rest of the system, 
\begin{equation}
\mathcal{C}_{re}(\rho^{A_{1}}) = Tr[\rho^{A_{1}} \log_{2}\rho^{A_{1}} - \rho^{A_{1}} \log_{2}\rho^{A_{1}}_{diag}]
\end{equation}
is the relative entropy of coherence, related to the visibility, with $\rho^{k}_{diag} = \sum_{i_{k}} \rho_{i_{k} i_{k}}^{k} \ket{i_{k}}\bra{i_{k}}$, and
\begin{equation}
\mathcal{P}_{vn}(\rho^{k}) = \log_{2}d_{k} - \mathcal{S}_{vn}(\rho^{k})
\end{equation}
is a measure of predictability.

For a system in a mixed state $\rho$, the CCR can be generalized to \cite{R12}
\begin{equation}
\label{CCRMISCELA}
\mathcal{S}_{k|B}(\rho)+\mathcal{P}_{vn}(\rho^{k})+\mathcal{C}_{re}(\rho^{k})+\mathcal{I}_{k:B}(\rho) = \log_{2}d_{k}
\end{equation}
where $\mathcal{I}_{k:B}(\rho)=\mathcal{S}_{vn}(\rho^{k}) + \mathcal{S}_{vn}(\rho^{B}) - \mathcal{S}_{vn}(\rho)$ is called mutual information, which is related to the entanglement between subsystem $k$ and $B=1 \cup 2 \cup k-1 \cup k+1 \cup... \cup n$, and $ \mathcal{S}_{k|B}(\rho) = \mathcal{S}_{vn}(\rho)-\mathcal{S}_{vn}(\rho^{B})$ is such that $\mathcal{S}_{k>B}(\rho) = - \mathcal{S}_{k|B}(\rho)$. This last quantity encodes the lack of knowledge about a part of the system when compared to the system as a whole.

\subsection{Chiral and flavor oscillations in neutrino time evolution}

In relativistic quantum mechanics the free evolution of a massive fermion is described by a wave function $\Psi$ whose evolution is given by the Dirac equation (in natural units $\hbar=c=1$)
\begin{equation}
(i\hat{\gamma}^{\mu}\partial_{\mu} - m)\Psi(x)=0 \label{eq:dirac}
\end{equation}
where the $\hat{\gamma}^{\mu}$ matrices satisfying the anticommutation relations
$\{ \hat{\gamma}^{\mu},\hat{\gamma}^{\nu} \} = 2\eta^{\mu \nu}$
with $\eta^{\mu \nu}= {\rm{diag}} \{ 1,-1,-1,-1 \}$. 
The solution $\Psi(x)$ is a $4-$component spinor, called a Dirac bispinor, which has two intrinsic DoFs: spin and chirality. From a group theory point of view this is due to the fact that the Dirac bispinors belong to irreducible representations of the \textit{complete} Lorentz group \cite{R19}.

Weak processes involve projections into states with definite chirality. For instance, in processes producing neutrinos and antineutrinos, these particles are always produced in states with a defined chirality. The chiral operator is defined by
\begin{equation}
\hat{\gamma}_{5} \equiv i\hat{\gamma}_{0}\hat{\gamma}_{1}\hat{\gamma}_{2}\gamma_{3},
\end{equation}
and we call its expectation value $\bra{\Psi}\hat{\gamma}^{5}\ket{\Psi}$ chirality. The chiral operator does not commute with the mass term of the free-particle Dirac equation, Eq.~(\ref{eq:dirac}), and therefore it is not a conserved quantity. Thus, a state that has an initial definite chirality will undergo chiral oscillations. Chiral oscillations for Dirac plane waves and wave-packets have been described, for example in \cite{A01,A02}. 

Chirality can be related, for massless particles, to the quantity called helicity, the projection of the spin along the direction of the momentum. The helicity operator is given by $\textbf{p} \cdot \hat{\mathbf{\Sigma}}/p $, where, in the standard representation of the Dirac matrices, $\hat{\mathbf{\Sigma}}={\rm{diag}}\{ \hat{\mathbf{\sigma}}, \hat{\mathbf{\sigma}}\}$. Since the helicity operator commutes with free-particle Dirac Hamiltonian (for massive and massless particles), it is often convenient to construct eigenspinors of the Dirac equation with definite helicity. The helicity operator commutes with the chiral operator, and for massless Dirac particles, the chiral operator also commute with the Dirac Hamiltonian. In this case, eigenspinors of the Dirac equation have both definite helicity and chirality, which indeed coincide. For massive particles, eigenspinors of the Dirac equation can have definite helicity, but not definite chirality. In this case, a state with definite chirality is necessarily a superposition of positive and negative energy eigenspinors.

The free-particle dynamics of the Dirac equation can be further extended to accommodate flavor-mixing and describe oscillating Dirac neutrinos. Within this framework, it can be shown that the free time evolution for an initial  electron neutrino state with left chirality  and  momentum $\textbf{p}$, is given by \cite{R14,R17} 
\begin{eqnarray}
\label{neutrino_time_evolution}
\ket{\nu_{e}(t)} &=& \big( \cos^{2}{\theta} \ket{\Psi_{m_{1}}(t)} + \sin^{2}{\theta} \ket{\Psi_{m_{2}}(t)} \big) \otimes \ket{1_{\nu_{e}}} \otimes \ket{0_{\nu_{\mu}}} \nonumber \\
&& + \sin{\theta} \cos{\theta} \big( \ket{\Psi_{m_{1}}(t)} - \ket{\Psi_{m_{2}}(t)} \big) \otimes \ket{0_{\nu_{e}} }\otimes \ket{1_{\nu_{\mu}}}.
\end{eqnarray}
where $\theta$ is the mixing angle and we used the flavor-qubit correspondence $\ket{\nu_e}\equiv \ket{1_{\nu_{e}}} \otimes \ket{0_{\nu_{\mu}}}$ and $\ket{\nu_\mu}\equiv \ket{0_{\nu_{e}} }\otimes \ket{1_{\nu_{\mu}}}$. The neutrino masses are indicated by $m_{1,2}$.

The time-evolved Dirac bispinor $\ket{\Psi_{m_{i}}(t)}$ is explicitly given by
\begin{equation}
\label{evolution_in_time_of_psi(0)}
\ket{\Psi_{m_{i}}(t)}=N_{i}\left(f_{+}e^{-iE_{i}t}\ket{u_{-}(\textbf{p},m_{i})}-f_{-}e^{iE_{i}t}\ket{v_{-}(-\textbf{p},m_{i})}\right), \qquad i=1,2
\end{equation}
with the Dirac eigenspinors
\begin{eqnarray}
\label{bispinori_uv}
\ket{u_{\pm}(\textbf{p},m_{i})} &=& N_{i}
\left( 
\begin{array}{c}
f_{\pm}^{i}\ket{\pm} \\
f_{\mp}^{i}\ket{\pm}
\end{array}
\right), \qquad \nonumber \\
\ket{v_{\pm}(\textbf{p},m_{i})} &=& N_{i}
\left( 
\begin{array}{c}
f_{\pm}^{i}\ket{\pm} \\
-f_{\mp}^{i}\ket{\pm}
\end{array}
\right),
\end{eqnarray}
and
\begin{equation}
\label{fattori f+-}
f_{\pm}^{i} \equiv f_{\pm}(\textbf{p},m_{i})=1\pm \frac{p}{E_{i}+m_{i}}, \qquad N_{i} \equiv N(\textbf{p},m_{i})=\sqrt{\frac{E_{i}+m_{i}}{4E_{i}}}.
\end{equation}

\vspace{0.2cm}

The state in Eq. (\ref{neutrino_time_evolution}) has three DoFs: chirality, spin and flavor. The flavor dynamics is obtained by tracing out the chirality and the spin from the density matrix $\rho_{e}(t)= \ket{\nu_{e}(t)} \bra{\nu_{e}(t)}$ which describes the joint spin-chirality-flavor state. The state given in Eq.~(\ref{neutrino_time_evolution}) can be written as
\begin{equation}
\label{eq:timeevolv}
\ket{\nu_{e}(t)}=\displaystyle\sum_{c=\pm1} \ket{C_{c}} \otimes \ket{\downarrow} \otimes \ket{\tilde{\Psi}(c)},
\end{equation}
where $\ket{\downarrow} \equiv \ket{-}$, $\ket{C_{+1(-1)}}$ is the left (right) chiral state and 
the flavor part of the state $\ket{\tilde{\Psi}(c)}$ reads
\begin{equation}
\ket{\tilde{\Psi}(c)} \equiv \delta_{e}(c) \ket{1_{\nu_{e}}} \ket{0_{\nu_{\mu}}} + \delta_{\mu}(c) \ket{0_{\nu_{e}}} \ket{1_{\nu_{\mu}}}, 
\end{equation}
with the time-dependent coefficients
\begin{eqnarray}
\delta_{e}(c) &\equiv& \cos^{2}\theta \,\omega_{1}(c) + \sin^{2}\theta\,\omega_{2}(c), \label{eq:delta_e} \\
\delta_{\mu}(c) &\equiv& \frac{\sin(2\theta)}{2}\big(\omega_{1}(c)-\omega_{2}(c) \big). \label{eq:delta_mu}
\end{eqnarray}
 and $\omega_{i}(c)$ is the probability amplitude for the chirality $c$ of the mass eigenstate $i$
\begin{eqnarray}
\omega_{i}(-1) &=& -i\frac{m_{i}}{E_{i}}\sin{(E_{i}t)}, \label{eq:omega_minus} \\
\omega_{i}(+1) &=& \cos(E_{i}t) - i\frac{p}{E_{i}}\sin(E_{i}t). \label{eq:omega_plus}
\end{eqnarray}

The time-evolved state in Eq.~(\ref{eq:timeevolv}) is a \textit{hyperentangled state} \cite{R20} comprising four qubits. Since the Dirac equation conserve helicity, the spin DoF remains separable through the time-evolution of the state, while quantum oscillations in the chirality and flavor sectors yield correlations between these partitions. Thus, chiral oscillations and the trade-off between quantum correlations in the different partitions of the system can have an effect on the flavor entanglement encoded in the state.

The density operator reduced to the flavor DoF reads
\begin{eqnarray}
\label{3.94}
\rho_{e\mu} &=& \sum_{c=\pm 1}\ket{\tilde{\Psi}(c)}\bra{\tilde{\Psi}(c)} \nonumber \\
&=& (\rho_{e \mu})_{11}(t) \ket{1_{e} 0_\mu}\bra{1_{e} 0_\mu} + (\rho_{e \mu})_{22}(t) \ket{0_{e} 1_\mu}\bra{0_{e} 1_\mu} \nonumber \\
&& + (\rho_{e \mu})_{12}(t) \ket{1_{e} 0_\mu}\bra{0_{e} 1_\mu} + (\rho_{e \mu})_{12}^*(t) \ket{0_{e} 1_\mu}\bra{1_{e} 0_\mu}.
\end{eqnarray}

The coefficients of the density matrix are given by
\begin{eqnarray}
\label{elementi matrice rho e mu}
(\rho_{e \mu})_{11}(t) &=& \sum_{c=\pm 1} \vert \delta_e (c) \vert^2 =1-\frac{\sin^{2}(2\theta)}{2}G(t), \nonumber \\
(\rho_{e \mu})_{12}(t) &=& \sum_{c=\pm 1}  \delta_e (c)  \delta_\mu^* (c)  =\frac{\sin(2\theta)\cos(2\theta)}{2}G(t)+i\frac{\sin(2\theta)}{2}H(t),
\end{eqnarray}

where we have defined the functions
\begin{eqnarray}
\label{definitions G H}
G(t) &=&  1- \left(\frac{p^{2}+m_{1}m_{2}}{E_{1}E_{2}} \right) \sin(E_{1}t)\sin(E_{2}t) - \cos(E_{1}t)\cos(E_{2}t), \nonumber \\
H(t) &=& \frac{p}{E_{1}}\sin(E_{1}t)\cos(E_{2}t) - \frac{p}{E_{2}}\cos(E_{1}t)\sin(E_{2}t).
\end{eqnarray}

The purity of the flavor-reduced state is given by
\begin{eqnarray}
\label{4.74}
Tr[\rho_{e\mu}^{2}] &=& |(\rho_{e \mu})_{11}|^{2}+|(\rho_{e \mu})_{22}|^{2}+2|(\rho_{e \mu})_{12}|^{2} \nonumber \\
&=& 1+\frac{\sin^{2}(2\theta)}{2} \left( G^{2}(t)+H^{2}(t)-2G(t) \right).
\end{eqnarray}

\begin{figure}[h!]
\centering
\subfloat[][]{\includegraphics[scale=0.63]{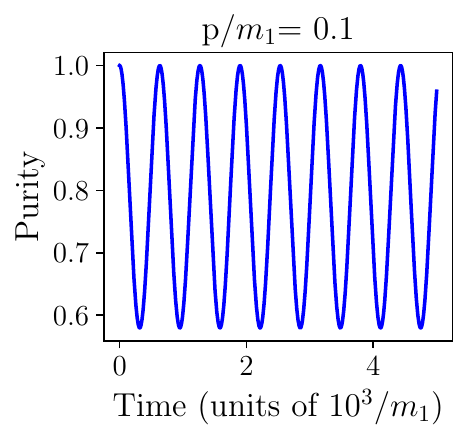}}
\subfloat[][]{\includegraphics[scale=0.63]{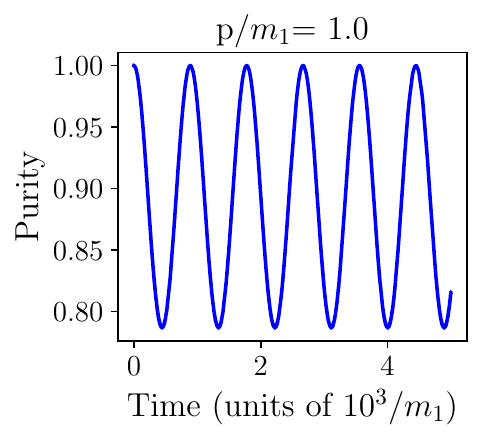}} 
\subfloat[][]{ \includegraphics[scale=0.63]{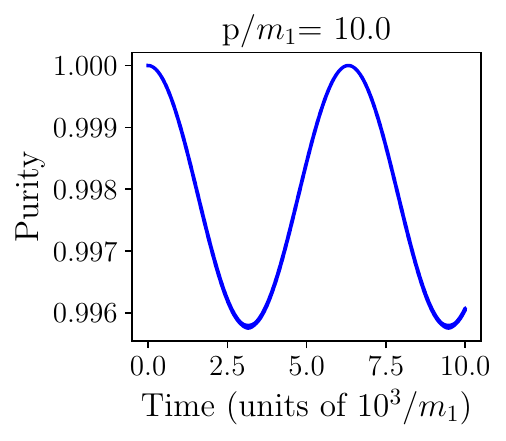}}
\caption{Flavor purity $Tr[\rho_{e \mu}^2]$ as a function of time for different values of the ratio $\frac{p}{m_{1}}$, with $\sin^{2}\theta=0.306$ and $\frac{\Delta{m^{2}}}{m_{1}}=0.001$ where $\Delta{m^{2}}=m_{2}^{2}-m_{1}^{2}$. For $p \gg m_{1}$ the amplitude of the oscillations in $Tr[\rho_{e\mu}^{2}]$ are less prominent, since chiral oscillations become less relevant and hence the flavor and chiral DoFs are disentangled.} \label{figrho2}
\end{figure}

In the relativistic limit $p \gg m_{1},m_{2}$ we obtain $G^{2}(t)+H^{2}(t)\simeq 2G(t)$ and hence $Tr(\rho_{e\mu}^{2}) \simeq 1$, which means that the subsystem is in a pure state and hence the total state described by $\rho$ does not exhibit entanglement between flavor and chirality, although there is still flavor entanglement. If we take the opposite limit $m_{1}, m_{2} \gg p$ we have that $G^{2}(t)+H^{2}(t)-2G(t) \simeq -\sin^{2}(\Delta{E}t)$ and hence 
\begin{equation}
Tr[\rho_{e\mu}^{2}] \simeq 1-\frac{\sin^{2}(2\theta)}{2}\sin^{2}(\Delta E t),
\end{equation}
with $\Delta{E}=E_{2}-E_{1}$. The purity in Eq.~(\ref{4.74} ) is useful to study the entanglement between the flavor and the chiral DoF: since the total system described by the state $\vert \nu_e (t) \rangle$, given in Eq.~(\ref{eq:timeevolv}), is pure, $Tr[\rho^{2}_{e\mu}]<1$ signals the presence of entanglement. We show the purity as a function of time in Fig.\ref{figrho2} for different values of the ratio $\frac{p}{m_{1}}$. A higher amplitude of the purity oscillations is obtained for smaller values of $\frac{p}{m_1}$, i.e. for non-relativistic particles exhibit a stronger chiral-flavor entanglement. The purity temporal evolution also displays higher frequency oscillations, as expected from standard chiral oscillations \cite{A01,A02}.

We can also obtain the density operators for the single flavors $e$ and $\mu$ by tracing out Eq.~(\ref{3.94}) with respect to one of the flavor modes. We obtain
\begin{eqnarray}
\rho_{e}=Tr_\mu[\rho_{e\mu}]=\rho_{11}\ket{1_{e}}\bra{1_{e}}+\rho_{22}\ket{0_{e}}\bra{0_{e}}, \\[2mm]
\rho_{\mu}=Tr_e[\rho_{e\mu}]=\rho_{22}\ket{1_{\mu}}\bra{1_{\mu}}+\rho_{11}\ket{0_{\mu}}\bra{0_{\mu}}
\end{eqnarray}

Hence the probability that the electronic neutrino does not change its flavor is
\begin{equation}\label{OscChiralavereged}
P_{ee}(t)=\rho_{11}=1-\frac{\sin^{2}(2\theta)}{2}G(t)
\end{equation}
Note that the last formula is not the same of the oscillation formula obtained in Ref. \cite{R17}, which accounts for both chiral and flavor oscillations. Indeed, the formula in Eq.~(\ref{OscChiralavereged}) is obtained by tracing on the chiral DoFs and in fact represents an upper bound to the formula of Ref.\cite{R17}. Note also that both these oscillation formulas contain the  non-standard oscillation term   obtained in a quantum field description of flavor mixing \cite{bla95,bla98}.

In the relativistic approximation $p \gg m_{1},m_{2}$, we can write $G(t) \simeq 1-\cos(\Delta Et)=2\sin^{2}\big( \frac{\Delta E t}{2} \big)$ and hence the probability reduces to
\begin{equation}
P_{ee}(t) \simeq P_{ee}^{S}(t) = 1-  \sin^{2}(2\theta)\sin^{2}\bigg( \frac{\Delta E t}{2} \bigg)
\end{equation}
that is the standard flavor oscillation formula. The same considerations hold for the transition probability $P_{e\mu}$.

\vspace{0,2cm}

We can now calculate the quantities involved in the CCR. In this case, we are interested in how different correlations encoded in the flavor DoFs are modified by chiral oscillations. As such, we consider a partition of the system between the flavor part (as a whole) and the the chirality and the spin part (as a whole). Since the total system is in a pure state, we use Eq.~(\ref{CCRPURO})
\begin{equation}
C_{re}(\rho_{e\mu})+P_{vN}(\rho_{e\mu})+S_{vN}(\rho_{e\mu})=\log_{2}(d_{e\mu}),
\end{equation}
and since we have a two qubit system, $d_{e\mu}=4$.

From Eq.~(\ref{elementi matrice rho e mu}) and the definition of the von Neumann entropy we obtain
\begin{equation}
S_{VN}(\rho_{e\mu}^{diag})=-\rho_{11}\log(\rho_{11})-\rho_{22}\log(\rho_{22}),
\end{equation}
and 
\begin{equation}
S_{VN}(\rho_{e\mu})=-\beta_{+}\log(\beta_{+})-\beta_{-}\log(\beta_{-}),
\end{equation}
where 
\begin{equation}
\beta_{\pm}= \frac{1\pm \sqrt{2 Tr[\rho_{e\mu}^{2}]-1}}{2}
\end{equation}
are the eigenvalues of flavor density operator $\rho_{e\mu}$.

\begin{figure}[t]
\centering
\subfloat[][]{\label{CCR1} 
\includegraphics[scale=0.65]{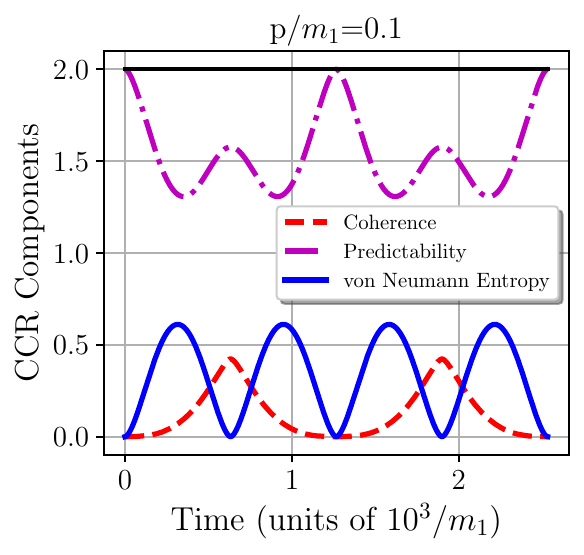}}
\subfloat[][]{\label{CCR2}\includegraphics[scale=0.65]{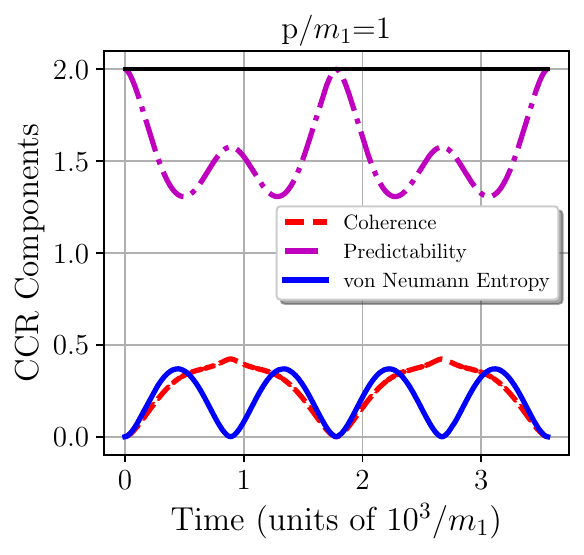}}\\
\subfloat[][]{\label{CCR3} \includegraphics[scale=0.65]{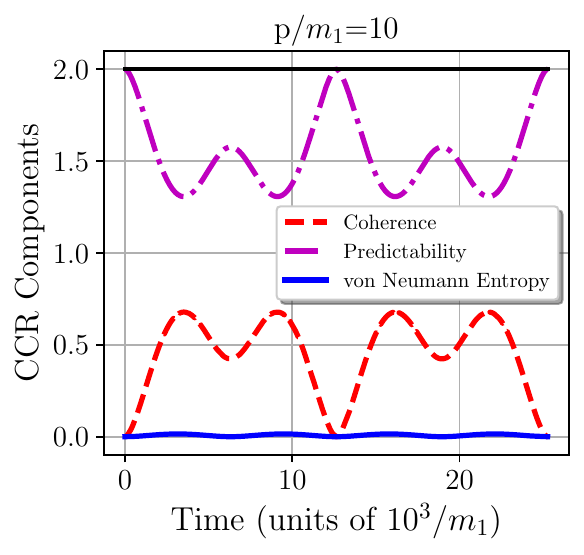}}
\caption{ Complete complementarity relations components (coherence, predictability and von Neumann entropy) as a function of the time respectively for different values of the ratio $\frac{p}{m_{1}}$, with $\sin^{2}\theta=0.306$ and $\frac{\Delta{m^{2}}}{m_{1}}=0.001$. For $p \gg m_{1}$ the von Neumann entropy vanishes.}
\end{figure}

We show in Figs.\ref{CCR1}, \ref{CCR2}, \ref{CCR3} the three quantities involved in the CCR as functions of time for different values of $\frac{p}{m_{1}}$. We observe that at $t=0$ the state is a pure electronic neutrino exhibiting no entanglement, and hence the predictability is maximum, while the von Neumann entropy and the relative entropy of coherence are zero. During the time evolution, the predictability decreases while entanglement and coherence increase, their sum always being constant according to the CCR. In the limit $\frac{p}{m_{1}}\gg1$, the entanglement goes to zero and only predictability and coherence are significant, since in such a limit chiral oscillations are negligible and, thus, they do not cause a redistribution of quantum correlations. In a more realistic model the wave function of the neutrinos should be described by wave packets with spread in momentum instead of plane waves. Wave-packets would then induce a decay of the flavor entanglement, but we expect that, in the non-relativistic limit, chiral oscillations would still persist. We postpone such an analysis for a future work.

\begin{figure}[t]
\centering
\subfloat[][]{\label{Svn1} \includegraphics[scale=0.63]{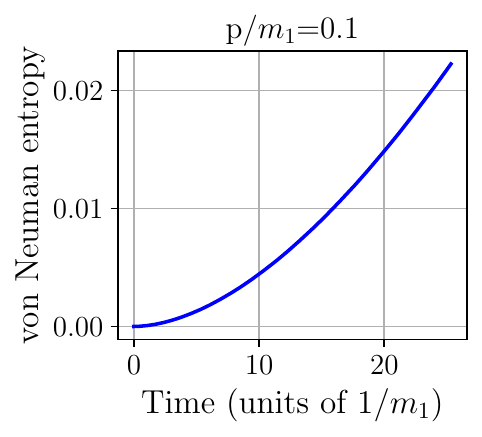}}
\subfloat[][]{\label{Svn2}\includegraphics[scale=0.63]{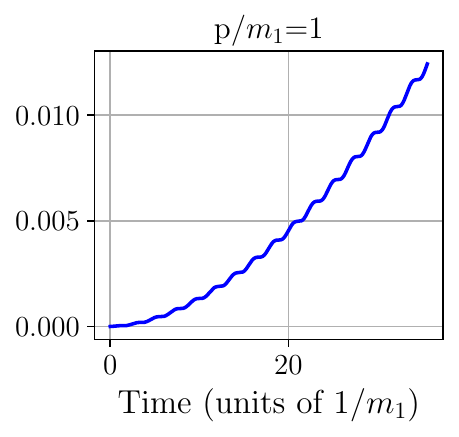}} 
\subfloat[][]{\label{Svn3} \includegraphics[scale=0.63]{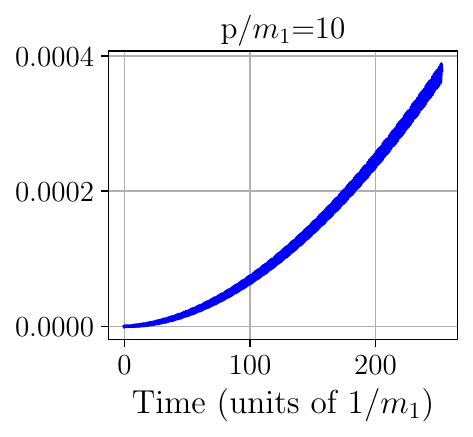}}
\caption{von Neumann entropy of the plots Fig.~\ref{CCR1}, \ref{CCR2} and \ref{CCR3} for different time scale showing the small oscillations, for the three different ratios $\frac{p}{m_{1}}$.}
\end{figure}

\section{CCR for a lepton-antineutrino pair}
\label{S3}
Weak interactions produce particles with a well definite chirality (left-chiral for leptons and right chiral for antileptons). Since each particle is initially in a well defined chiral states, the dynamics under the free Dirac equation will induce chiral oscillations. In \cite{R18}, it was shown that chiral oscillations can modify the spin correlations shared in a lepton-antineutrino pair produced by a pion decay.  Nevertheless, the antineutrino also undergoes flavor oscillations, which can affect the distribution of quantum correlations in other partitions of the system, as we described in the last section. We now characterize such interplay between quantum correlations in lepton-antineutrino pair by means of the CCRs.

As a simple example, we can consider the decay of a pion into the two channels:
\begin{equation}
\pi^{-} \rightarrow \mu^{-} + \bar{\nu}_{\mu}, \qquad \pi^{-} \rightarrow e^{-} + \bar{\nu}_{e}
\end{equation}
with probabilities $99,9877\%$ and $0,0123 \%$, respectively\footnote{The mismatch between helicity and chirality yields the suppression of the decay into the electron channel, which should be favored for kinematic reasons.}. In the rest frame of the pion, the two particles are emitted along opposite directions and with a well defined chirality: right-handed for the antineutrino and left-handed for the lepton. Furthermore, due to conservation of angular momentum, the total spin of the emitted particles has to vanish. Assuming  that the lepton with mass $m_{l}$ has a momentum $-p\hat{e}_{z}$ and the antineutrino  $p\hat{e}_{z}$,  we can take the initial state in the form
\begin{equation}
\label{phiZero}
\ket{\Phi}=\frac{\ket{v_{\uparrow}(p,m_{\bar{\nu}})} \otimes \ket{u_{\downarrow}(-p,m_{l})}-\ket{v_{\downarrow}(p,m_{\bar{\nu}})} \otimes \ket{u_{\uparrow}(-p,m_{l})}}{\sqrt{2}}
\end{equation}
which has entangled spins. To model the well definite chirality, we project the state in Eq.~(\ref{phiZero}) into chirality eigenstates 
\begin{equation}
\label{numeratore psi(0)}
\ket{\Psi(0)}=\frac{\hat{\Pi}_{R}^{\bar{\nu}} \otimes \hat{\Pi}_{L}^{l}{\ket{\Phi}}}{\langle \Phi \vert \hat{\Pi}_{R}^{\bar{\nu}} \otimes \hat{\Pi}_{L}^{l} \vert \Phi \rangle}
\end{equation}
where the chirality projectors are given by
\begin{equation}
\hat{\Pi}_{R/L}^{\alpha}=\frac{I^{\alpha}\pm\hat{\gamma}_{5}^{\alpha}}{2}
\end{equation}
with $\alpha=l,\bar{\nu}$.

After some algebraic manipulation and using the relation $p^{2}=E_{\bar{\nu}}^{2}-m_{\bar{\nu}}^{2}=E_{l}^{2}-m_{l}^{2}$, the time evolution of the state in Eq.~(\ref{numeratore psi(0)}) becomes \cite{R18}
\begin{equation}
\label{evoluzione temporale psi chirality e spin}
\ket{\Psi(t)}=A(p,m_{l},m_{\bar{\nu}})\ket{\psi_{\bar{\nu}_{\uparrow}}(t)} \otimes \ket{\psi_{l_{\downarrow}}(t)} - B(p,m_{l},m_{\bar{\nu}})\ket{\psi_{\bar{\nu}_{\downarrow}}(t)} \otimes \ket{\psi_{l_{\uparrow}}(t)},
\end{equation}
with
\begin{eqnarray}\label{evtempbisp1}
\ket{\psi_{\bar{\nu}_{\uparrow(\downarrow)}}(t)}&=&N_{\bar{\nu}}[e^{-iE_{\bar{\nu}}t}f_{+}^{\bar{\nu}}\ket{u_{\uparrow(\downarrow)}(p,m_{\bar{\nu}})}+e^{iE_{\bar{\nu}}t}f_{-}^{\bar{\nu}}\ket{v_{\uparrow(\downarrow)}(-p,m_{\bar{\nu}})}]
\\
\label{evtempbisp4}
\ket{\psi_{l_{\uparrow(\downarrow)}}(t)}&=&N_{l}[e^{-iE_{l}t}f_{+}^{l}\ket{u_{\uparrow(\downarrow)}(-p,m_{l})}-e^{iE_{l}t}f_{-}^{l}\ket{v_{\uparrow(\downarrow)}(p,m_{l})}]
\end{eqnarray}
and 
\begin{equation}
\label{Coefficienti A e B}
\begin{array}{l}
A(p,m_{l},m_{\bar{\nu}})=N_{l}N_{\bar{\nu}}f_{+}^{\bar{\nu}}f_{-}^{l}\bigg[ \frac{1}{2}-\frac{p^{2}}{2E_{l}E_{\bar{\nu}}} \bigg]^{-\frac{1}{2}}, \\[2mm]
B(p,m_{l},m_{\bar{\nu}})=N_{l}N_{\bar{\nu}}f_{-}^{\bar{\nu}}f_{+}^{l}\bigg[ \frac{1}{2}-\frac{p^{2}}{2E_{l}E_{\bar{\nu}}} \bigg]^{-\frac{1}{2}}.
\end{array}
\end{equation}

The coefficients $f_{\pm}^{\bar{\nu}(l)}$ are given by Eq.~(\ref{fattori f+-}). For a relativistic antineutrino $p\gg m_{\bar{\nu}}$ and $f_{-}^{\bar{\nu}} \rightarrow 0$. Hence $A \gg B$, and only the first term in Eq.~(\ref{evoluzione temporale psi chirality e spin}) contributes to the state.  Chiral oscillations will be relevant for the superposition in the opposite limit, in which $B$ is comparable to $A$. 

Flavor is another intrinsic DoF for the neutrino. Charged weak interactions produce neutrinos with a well defined flavor, which is not conserved during free evolution; akin to chiral oscillations, we expect that flavor oscillations will also redistribute the initial entanglement between spins. For simplicity we restrict our analysis to a two-flavors scenario (electronic neutrino $e$ and muonic neutrino $\mu$).

We then consider the initial state
\begin{equation}
\label{TempEvoFlav}
\begin{array}{l}
\fl \ket{\Psi(0)}=A(p,m_{l},m_{\bar{\nu}, 1})\ket{\psi_{\bar{\nu}_{e, \uparrow}}(0)} \otimes \ket{\psi_{l_{\downarrow}}(0)} 
- B(p,m_{l},m_{\bar{\nu}, 1})\ket{\psi_{\bar{\nu}_{e, \downarrow}}(0)} \otimes \ket{\psi_{l_{\uparrow}}(0)},
\end{array}
\end{equation}
where $\ket{\psi_{\bar{\nu}_{e, \uparrow}}(0)} $ is the state of an electron antineutrino with initially right-handed chirality, given by \cite{A01,R17,A02}
\begin{equation}
\ket{\psi_{\bar{\nu}_{\uparrow (\downarrow)}} (0)}= \vert \psi_{R,\uparrow(\downarrow)} (0) \rangle \otimes \vert \bar{\nu}_e \rangle,
\end{equation}
with $\vert \psi_{R,\uparrow(\downarrow)}  (0) \rangle$ a right-handed bispinor with the corresponding spin polarization. We use this phenomenological description since, in the limit of no flavor oscillation, one expects the electron neutrino to have the lightest mass. Such an initial state corresponds to Eq.~(\ref{evoluzione temporale psi chirality e spin}) for the mass eigenstate $1$ projected into the electron neutrino flavor.

The temporal evolution of such flavor state reads \cite{R14,A01,R17,A02}
\begin{equation}
\label{eq:antineutrinotevol}
\begin{array}{l}
\ket{\psi_{\bar{\nu}_{\uparrow(\downarrow)}}(t)} = \bigg( \cos^{2}\theta \ket{\psi_{\bar{\nu}_{1\uparrow(\downarrow)}}(t)} + \sin^{2}\theta \ket{\psi_{\bar{\nu}_{2\uparrow(\downarrow)}}(t)} \bigg) \otimes \ket{\bar{\nu}_{e}} \\[2mm]
+\sin\theta \cos\theta \bigg( \ket{\psi_{\bar{\nu}_{1\uparrow(\downarrow)}}(t)}-\ket{\psi_{\bar{\nu}_{2\uparrow(\downarrow)}}(t)} \bigg) \otimes \ket{\bar{\nu}_{\mu}},
\end{array}
\end{equation}
where $\theta$ is the mixing angle, and the time evolution of the mass bispinors $\ket{\psi_{\bar{\nu}_{i\uparrow(\downarrow)}}}$ is given as in Eqs. (\ref{evtempbisp1}) where the index $i$ is related to the mass $m_{i}$, while the states $\ket{\bar{\nu}_{e}} = \ket{1_{e}0_{\mu}}$ and $\ket{\bar{\nu}_{\mu}} = \ket{0_{e}1_{\mu}}$ describe the flavor DoFs.

The time evolved state is given by using Eq.~(\ref{eq:antineutrinotevol}) and Eqs.~(\ref{evtempbisp1}), which after some algebraic steps gives
\begin{eqnarray}
\label{eq:fullstate}
\ket{\Psi(t)} &=& \sum_{c,c'} \bigg(  \gamma_{1}(c,c')\ket{C_{\bar{\nu}}^{c}} \otimes \ket{\uparrow_{\bar{\nu}}} \otimes \ket{\bar{\nu}_{e}} \otimes \ket{C_{l}^{c'}} \otimes \ket{\downarrow_{l}} \nonumber \\
&& - \gamma_{2}(c,c')\ket{C_{\bar{\nu}}^{c}} \otimes \ket{\uparrow_{\bar{\nu}}} \otimes \ket{\bar{\nu}_{\mu}} \otimes \ket{C_{l}^{c'}} \otimes \ket{\downarrow_{l}} \nonumber \\
&& - \gamma_{3}(c,c')\ket{C_{\bar{\nu}}^{c}} \otimes \ket{\downarrow_{\bar{\nu}}} \otimes \ket{\bar{\nu}_{e}} \otimes \ket{C_{l}^{c'}} \otimes \ket{\uparrow_{l}} \nonumber \\
&& - \gamma_{4}(c,c')\ket{C_{\bar{\nu}}^{c}} \otimes \ket{\downarrow_{\bar{\nu}}} \otimes \ket{\bar{\nu}_{\mu}} \otimes \ket{C_{l}^{c'}} \otimes \ket{\uparrow_{l}} \bigg). \label{psitflavor}
\end{eqnarray}

with $c,c'=\pm 1$. The coefficients $\gamma_i (c, c^\prime)$ are given in the Appendix.

To compute the quantities involved in the CCR for the spin part (as a whole) and the rest of the system, we trace out the chirality and the flavor DoFs. The procedure yields the reduced density operator in the basis $\{ \vert \uparrow_{\bar{\nu}} \downarrow_{l} \rangle, \vert \downarrow_{\bar{\nu}} \uparrow_{l} \rangle \}$,
\begin{equation}
\label{density operator SpinChiralityflavor}
\rho_{S_{\bar{\nu}},S_l}(t) = \left(
\begin{array}{cc}
A^{2} & \rho_{12} \\[2mm]
\rho_{12}^{*} & B^{2}
\end{array}
\right),
\end{equation}
where the coefficients $A$ and $B$ are given in Eqs.~(\ref{Coefficienti A e B}), and
\begin{equation}
\begin{array}{l}
\rho_{12} = AB \Big\{ \cos^{4} \theta ( h_{l} h_{\bar{\nu}_{1}} + g_{l} g_{\bar{\nu}_{1}} ) + \sin^{4} \theta ( h_{l} h_{\bar{\nu}_{2}} + g_{l} g_{\bar{\nu}_{2}} ) \\ 
+ \sin^{2} \theta \cos^{2} \theta [ h_{l} ( h_{\bar{\nu}_{1}} + h_{\bar{\nu}_{2}} ) + g_{l} ( g_{\bar{\nu}_{1}} + g_{\bar{\nu}_{2}} ) ] \\
+ i \cos^{4} \theta ( g_{l} h_{\bar{\nu}_{1}} - g_{\bar{\nu}_{1}} h_{l} ) + i \sin^{4} \theta ( g_{l} h_{\bar{\nu}_{2}} - g_{\bar{\nu}_{2}} h_{l} ) \\
+ i \sin^{2} \theta \cos^{2} \theta [ g_{l} ( h_{\bar{\nu}_{1}} + h_{\bar{\nu}_{2}} ) - h_{l} ( g_{\bar{\nu}_{1}} + g_{\bar{\nu}_{2}} ) ] \Big\}
\end{array}
\end{equation}
with
\begin{equation}
h_{i}=1-\frac{2p^{2}}{E_{i}^{2}}\sin^{2}(E_{i}t)
,\quad 
g_{i}=\frac{p}{E_{i}}\sin(2E_{i}t)
\end{equation}
for $i=l$, $\bar{\nu}_{1}$, $\bar{\nu}_{2}$. The purity of the spin-reduced density matrix is then given by
\begin{eqnarray}\nonumber
\fl Tr[\rho^{2}_{S_{\bar{\nu},S_l}}(t)] &= 1-\frac{\mathcal{N}[\rho_{S_{\bar{\nu},S_l}}^{2}(0)]}{2} \bigg\{ 1-\Gamma_{l}^{2} \Big( \cos^{8} \theta \Gamma_{\bar{\nu}_{1}}^{2} + \sin^{8} \theta \Gamma_{\bar{\nu}_{2}}^{2} \Big) \\
& -2 \Gamma_{l}^{2}  \sin^{2} \theta \cos^{2} \theta \Big[ 2 \cos^{4}\theta \Gamma_{\bar{\nu}_{1}}^{2}+ 2 \sin^{4}\theta \Gamma_{\bar{\nu}_{1}}^{2}+ \sin^{2}\theta \cos^{2}\theta (\Gamma_{\bar{\nu}_{1}}^{2} + \Gamma_{\bar{\nu}_{2}}^{2}) \nonumber \\
\label{Traccia rho quadro spin chirality flavor caso generale}
&\hspace{3.5 cm}+ 2(h_{\bar{\nu}_{1}} h_{\bar{\nu}_{2}} + g_{\bar{\nu}_{1}} g_{\bar{\nu}_{2}}) \Big] \bigg \}, 
\end{eqnarray}
with
\begin{equation*}
\Gamma_{i} = h_{i}^{2}+g_{i}^{2},
\end{equation*}
and $i=l,\bar{\nu}_{1},\bar{\nu}_{2}$. The logarithmic negativity at $t=0$ is given by
\begin{equation}
\label{ent_t=0}
\mathcal{N}[\rho_{S_{l},S_{\bar{\nu}}}(0)]=2|AB|=2|A(p,m_{l},m_{\bar{\nu}})B(p,m_{l},m_{\bar{\nu}})|.
\end{equation}
As a check, we observe that when flavor mixing is absent $(\theta=0)$, the above expression reduces to
\begin{equation}
Tr[\rho^{2}_{S_{\bar{\nu},S_l}}(t)]= 1-\frac{\mathcal{N}[\rho_{S_{\bar{\nu},S_l}}^{2}(0)]}{2} \bigg( 1-\Gamma_{l}^{2}   \Gamma_{\bar{\nu}_{1}}^{2} \bigg)
\end{equation}
which coincides with the result in \cite{R18} as we expected.

We can now evaluate the different components of the CCR in two cases. The first one is for the subsystem for which we evaluate the CCR formed by both spins and the second subsystem formed by all the other DoFs, chirality and flavor. In this case, using the CCR we will analyze how the correlations shared between the spins  evolve with respect to correlations shared with other parts of the system. Since the total system is in a pure state we use Eq.~ (\ref{CCRPURO}). The second case, corresponds to describing the spin part as composed by the spin of the electron and the spin of the antineutrino. In this case, we can focus on the state of one of the spins, which should be a mixed state. We use the CCR in Eq.~(\ref{CCRMISCELA}) to compute how chiral and flavor oscillations redistributes the entanglement and other correlations between the spins.

\vspace{0,2cm}

\begin{figure}[t]
\centering
\subfloat[][]{\label{CCR total system 1} \includegraphics[scale=0.45]{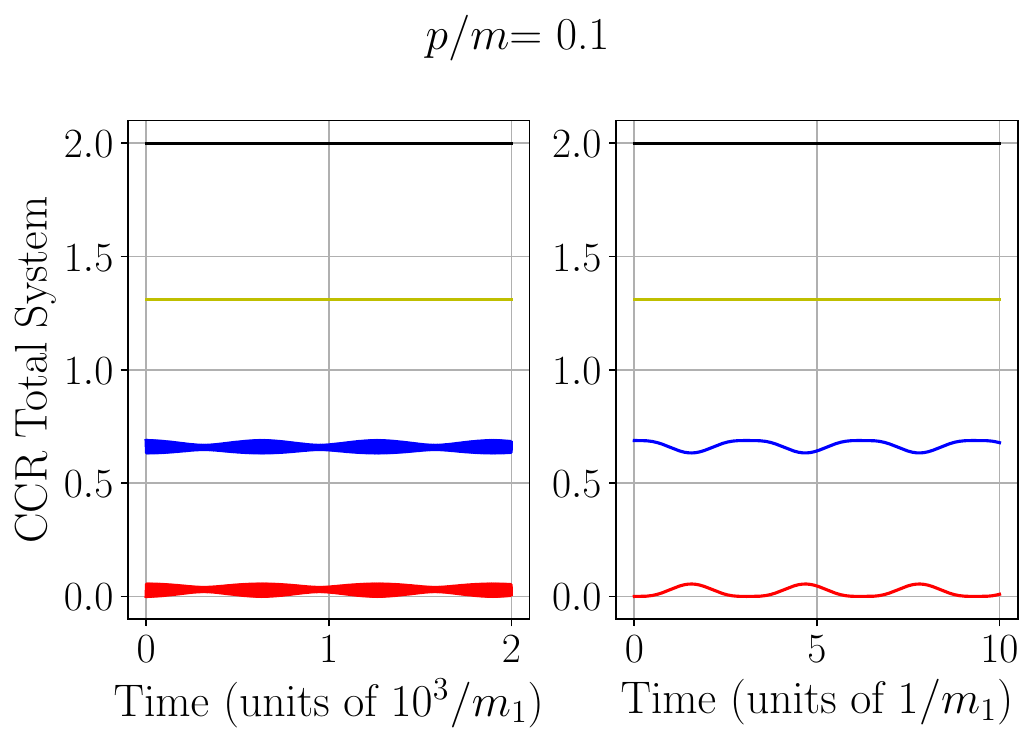}}
\subfloat[][]{\label{CCR total system 2} \includegraphics[scale=0.45]{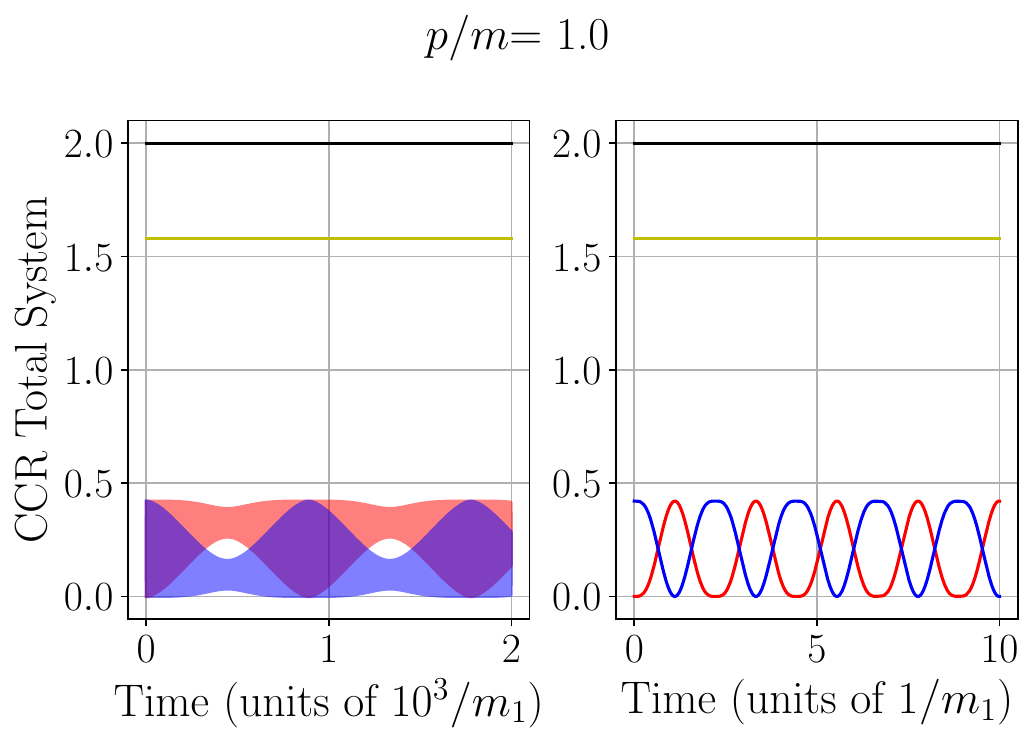}}\\
\subfloat[][]{\label{CCR total system 3} \includegraphics[scale=0.45]{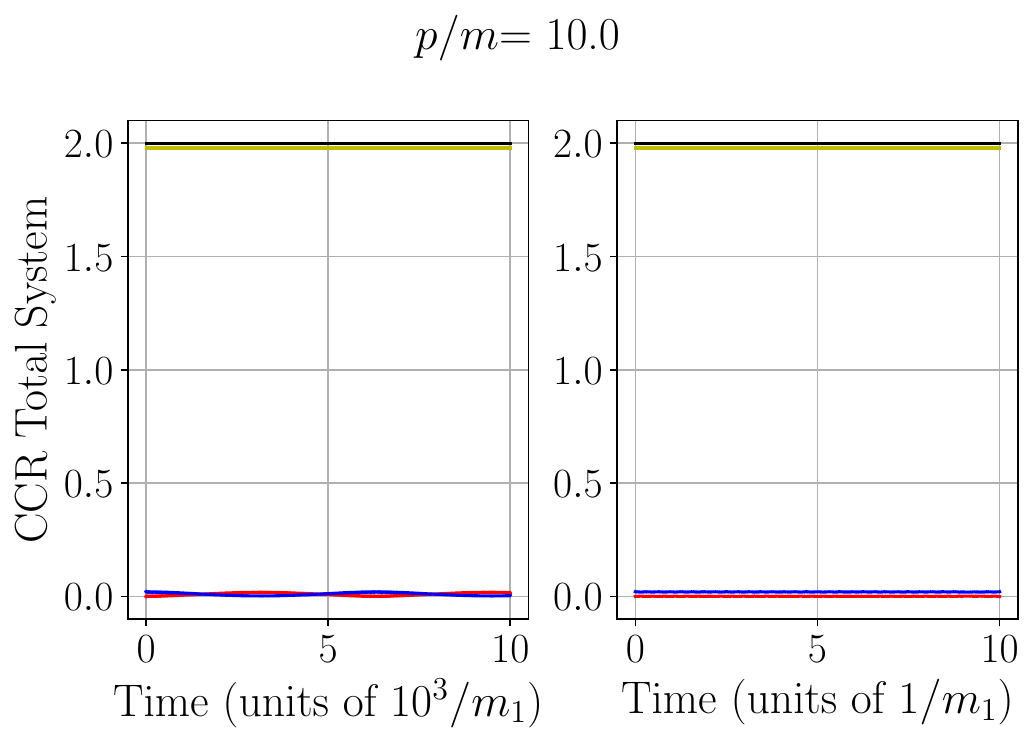}}
\caption{CCR components  as a function of time for $\frac{p}{m_{\bar{\nu}_1}}=0.1,1,10$ and values of the parameters as above: predictability (yellow), von Neumann entropy (red),  coherence (blue).} \label{Fig:04}
\end{figure}

In the first case, shown in Fig. \ref{Fig:04}, the predictability is constant in time, since it is related to the diagonal elements of the reduced density operator for the spin part, which are time-independent. Furthermore the predictability increases with the ratio $p/m_1$, because an increasing in this ratio yields an increase in the difference between the coefficients $A$ and $B$. The coherence decreases when the $p/m_{\bar{\nu}_1}$ increases, and the von Neumann entropy, related to the entanglement between the spin part and the rest of the DoFs, increases when the momentum is of the same order as the antineutrino mass. The effects of the chiral and the flavor oscillations involve only the last two quantities, and they become more relevant for $p\sim m_{\bar{\nu}_1}$.

\begin{figure}[h!]
\centering
\includegraphics[scale=0.8]{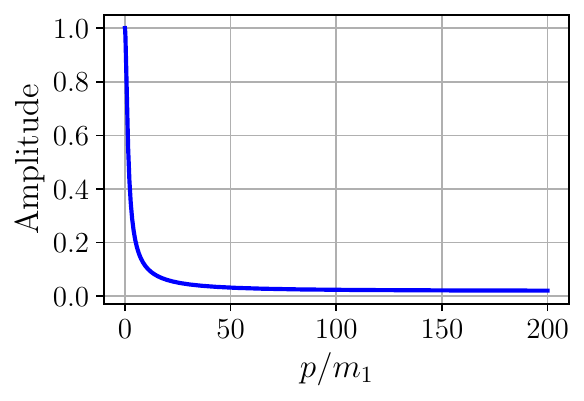}
\caption{Plot of the oscillation amplitude Eq.(\ref{ent_t=0}) in terms of the ratio $\frac{p}{m_{\nu}}$. Note that the amplitude does not vanish when $p \to \infty$ but it has asymptotic value $\frac{m_{\nu} m_{l}}{m_{\nu}^{2}+m_{l}^{2}}$, see Eq.~(\ref{eq:limitMT}).} \label{Fig_amplitude}
\end{figure}

Similarly, when we focus on the CCR for a single spin, the predictability is always constant while the coherence is zero. The mutual information related to the entanglement increases (decreases) in time while the mutual information decreases (increases). As in the previous case, such effects become more relevant in the limit $p \sim {m_{\bar{\nu}_1}}$. Note that the amplitude of the oscillations (\ref{ent_t=0}) also depends on the amount of entanglement between the spins at $t=0$. As we have already mentioned this quantity becomes small when $p \to \infty$, but it does not vanishes. We show such a behavior in Fig.~\ref{Fig_amplitude}, where we plot Eq.~(\ref{ent_t=0}) as a function of $p/m_{\bar{\nu}_1}$, for a fixed $m_l/m_{\bar{\nu}_1}$. Such an amplitude decays exponentially and for $p \gg m_{\bar{\nu}_1}, m_{l}$ tends to
\begin{equation}
\label{eq:limitMT}
  \lim_{p/m_{\bar{\nu}_1} \to \infty}  \mathcal{N}[\rho_{S_{l},S_{\bar{\nu_{e}}}}(0)] = \frac{m_{\nu} m_{l}}{m_{\nu}^{2}+m_{l}^{2}}.
\end{equation}
For the parameters considered in Fig.~\ref{Fig_amplitude}, such a limit turns out to be small.

\begin{figure}
\centering
\subfloat[][]{\label{CCR spin subsystem 1} \includegraphics[scale=0.45]{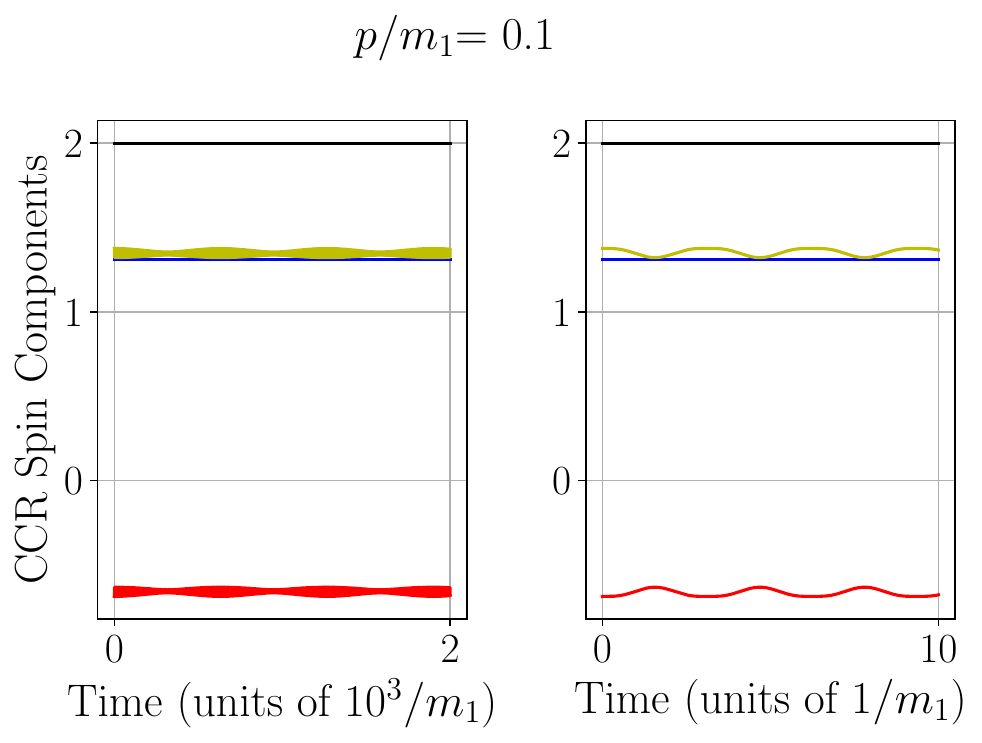}}
\subfloat[][]{\label{CCR spin subsystem 2} \includegraphics[scale=0.45]{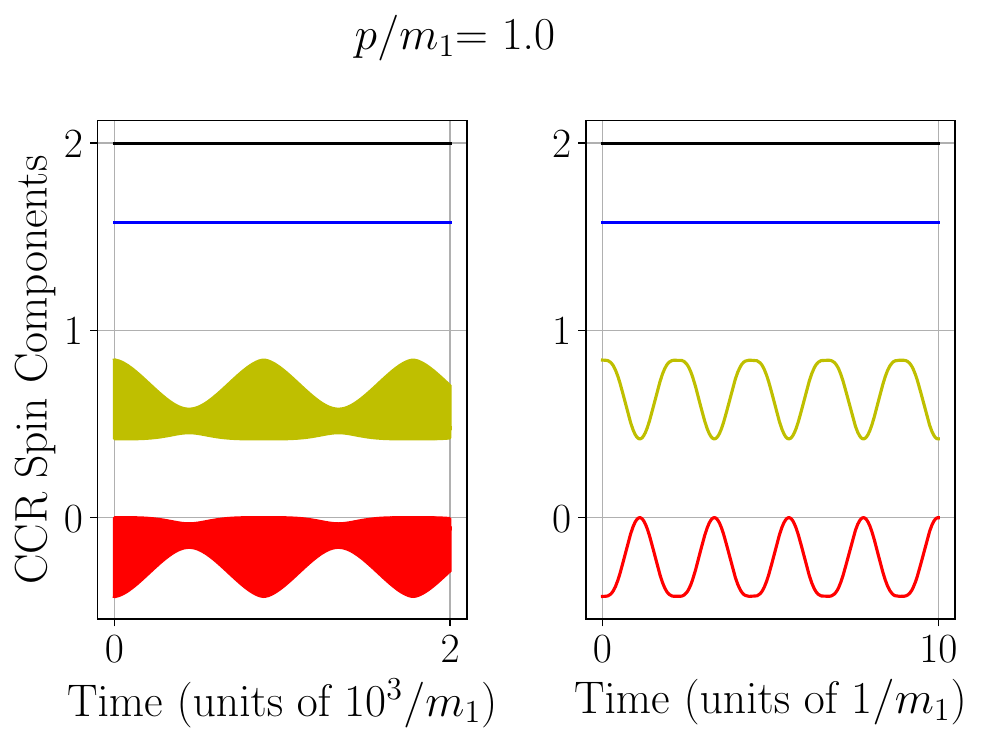}}\\
\subfloat[][]{\label{CCR spin subsystem 3}\includegraphics[scale=0.45]{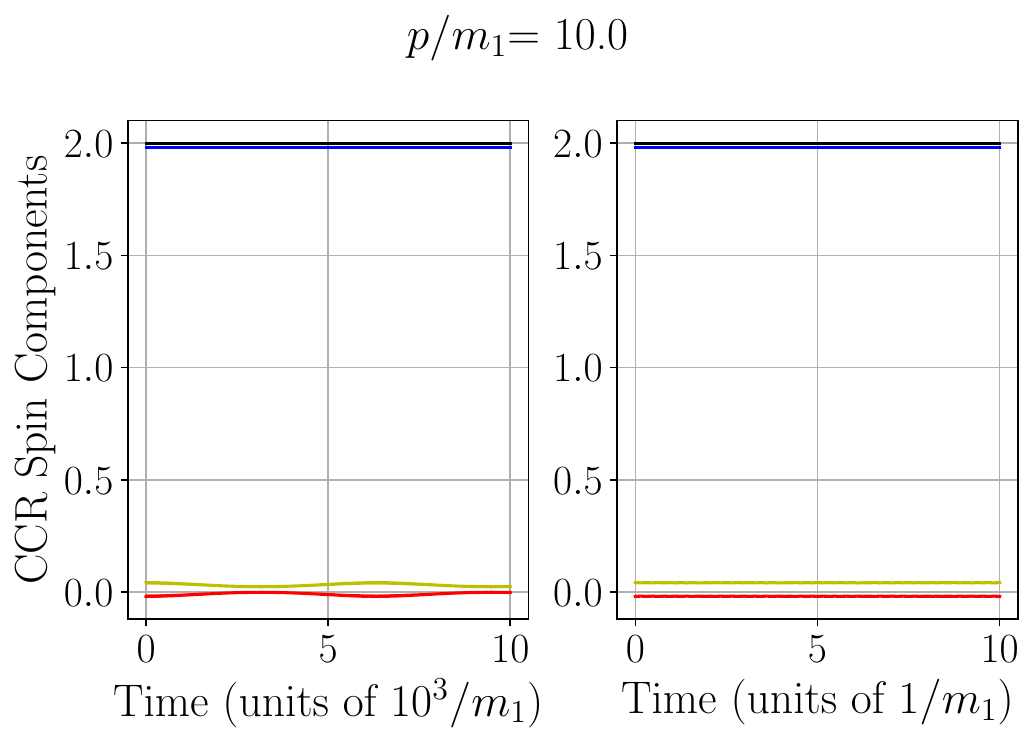}}
\caption{CCR are plotted as a function of time for $\frac{p}{m_{\bar{\nu}_1}}=0.1,1,10$ and values of the parameters as above: mutual information (yellow),  predictability (blue) and the loss of information of the subsystem (red) (see Eq. \ref{general_CCR_mixed}) .}
\end{figure}

\newpage
\section{Conclusions}

We have investigated the dynamics of quantum correlations encoded in Dirac neutrino states, and in lepton-antineutrino pairs, as a consequence of both chiral and flavor oscillations. The mass term of the Dirac equation mixes components of the bispinor associated with different chiralities. Since flavor mixing requires a superposition of mass eigenstates, there is an interplay between chiral and flavor oscillations which yields the generation of states in which all the intrinsic DoFs of a neutrino, chirality, spin and flavor, are entangled (hyperentanglement). We have characterized the interplay between the different correlations shared among the DoFs by means of CCRs. In particular, we have shown that there is a redistribution of entanglement and coherence in the flavor sector by the chiral oscillations. Finally, we considered the dynamics of an entangled lepton-antineutrino pair with an effective model in which each component of the pair initially exhibits a definite chirality. Using CCRs, we have shown that both chiral and flavor oscillations can affect correlations shared among the spins of the particles. In both scenarios, chiral oscillation effects are more prominent in non-relativistic dynamical regime. The amplitudes of the oscillations displayed by the quantum correlations peaks for momenta of the (neutrino) antineutrino comparable to its mass.

Our analysis can pave the way for potential investigations of chiral oscillations by means of spin entanglement measurements in lepton-antineutrino pairs, and to elucidate chiral oscillation effects in non-relativistic neutrinos. An example of the latter is the \textit{cosmic neutrino  background} $(C\nu B)$ \cite{Dey2024} with a temperature of $T \simeq 1,95K$. Although the $C\nu B$ has been never measured, several experiments have been designed and are currently being implemented to probe the $C\nu B$  \cite{ptolemy}, in which case chiral oscillations can be relevant \cite{R17,GePasquini}. Furthermore, non-relativistic neutrinos could also be measured in accelerator experiments \cite{Bauer2021}. In such a context, chiral oscillations can also be relevant for distinguishing Dirac from Majorana neutrinos \cite{R17, GePasquini, Hernandez2022,Li:2023iys}. Our analysis can be further use to assess whether neutrinos are useful for, e.g., implementing quantum communication protocols \cite{QcomNeut}. Finally, flavor oscillations are modified when neutrinos propagate in matter, \cite{Kuo1989}-\cite{E01}, which is  also an interesting framework to investigate the interplay between chiral and flavor oscillations. A further extension of our work includes a complete quantum field theoretical description of chiral and flavor oscillations, following the approach proposed in \cite{ArXiV}.

\section*{Acknowledgements}

V.A.S.V.B. acknowledges financial support by the Contrat Triennal 2021-2023 Strasbourg Capitale Europeenne.

\clearpage

\appendix

\section{Summary of Complete Complementarity Relations}
The \textit{complementary principle} is one of the fundamental concepts of quantum mechanics. The term was introduced by Niels Bohr \cite{R1} and it states that quantum systems have certain complementarity properties which cannot be observed or measured simultaneously. This happens because it is not always possible to make a measurement on a quantum system without perturbing it.

\vspace{0,2cm}

One of the features of complementarity principle is the ``wave-particle duality'' which states that a quantum object can behave like a wave or a particle depending on the experiment that one performs on the system: 
a situation in which these features are well represented is the double slit experiment. In this experiment, the interference pattern is observed only if the observer does not measure the path followed by the photons which compose the light. 

However, for many years the wave-particle duality was only a conceptual method to understand fundamental aspects of quantum mechanics and only in 1978 Wootters and Zurek \cite{R2} were able to obtain a quantitative formulation of this problem. 

They also give an alternative form of double slit experiment in which the two slits have different widths. In this way, there is a different probability that the photons pass throw them and hence it gives an a-priori knowledge on which way is taken by the photon. With a such apparatus, they showed the interference pattern decreases but does not disappear completely.

The problem was later studied \cite{R3}-\cite{R5} for a Michelson and Morley interferometer instead of a Young interferometer, but there is no conceptual difference between the two cases. By introducing a mechanism able to obtain a partial which-way information, they found an inequality
\begin{equation}
\label{disuguaglianza visibilità predicabilità}
V^{2}+P^{2} \leq 1
\end{equation}
where $V$ stands for \textit{visibility} and $P$ for \textit{predictability}. The visibility quantifies the interference pattern, and so it is related also to the coherence of the system, while the predictability quantifies the (a-priori) particle-like behaviour. The equality is fulfilled only by pure states. 
D\"urr generalized the inequality $(\ref{disuguaglianza visibilità predicabilità})$ to the case of a $n$ multi-beam interferometer \cite{R6}.

The above inequality gives a satisfying description only for pure states. Indeed  it can be shown that for mixed states both visibility and predictability can decrease (or increase) simultaneously \cite{R7}.

The apparent problem here is that the predictability is a measure of only a-priori particle-like behaviour, since no which-way detector has been taken into account. If a detector is included in the system, it can measure the path followed by the particle. Supposing that the detector is also a quantum object, it has to interact with the system to make the measurement and hence entanglement between the detector and the particle occurs.
This was noticed by Jakob and Bergou \cite{R_CCR_qubit} which generalized the inequality (\ref{disuguaglianza visibilità predicabilità}) for a generic pure bipartite two-qubit system, that is
\begin{equation}
\label{CCR_ineq}
\mathcal{P}_{k}^{2}+\mathcal{V}_{k}^{2}+\mathcal{C}^{2} \leq 1
\end{equation}
where $k=1,2$ stands for the two subsystems and $\mathcal{C}$ is the concurrence \cite{R_concurrence}, a measure of the entanglement between the two parts given by 
\begin{equation}
\mathcal{C} = \max \{0, \lambda_{1}-\lambda_{2}-\lambda_{3}-\lambda_{3} \}
\end{equation}
where the $\lambda_{i}$ are the eigenvalues (in decreasing order) of the operator $\rho \tilde{\rho}$ with 
\begin{equation}
\tilde{\rho} \equiv \sigma_{y} \otimes \sigma_{y} \rho^{*} \sigma_{y} \otimes \sigma_{y} 
\end{equation}
is the spin-flipped density operator $\rho$.
The concurrence is related to the entanglement of formation by \cite{R_concurrence}
\begin{equation}
\label{EoF}
\begin{array}{l}
EoF = -\frac{1+\sqrt{1-\mathcal{C}^{2}}}{2}\log_{2}{\bigg( \frac{1+\sqrt{1-\mathcal{C}^{2}}}{2}\bigg)} -\frac{1-\sqrt{1-\mathcal{C}^{2}}}{2}\log_{2}{\bigg( \frac{1-\sqrt{1-\mathcal{C}^{2}}}{2}\bigg)}
\end{array}
\end{equation}

If the total system is in a pure state, Eq.(\ref{CCR_ineq}) becomes an equality: 
\begin{equation}
\mathcal{P}_{k}^{2}+\mathcal{V}_{k}^{2}+\mathcal{C}^{2} = 1
\end{equation}
Notice that the equality is preserved also when entanglement occurs and the two subsystems are mixed. If the two subsystems does not interact with each other, both  are pure states and Eq.(\ref{CCR for pure qubit}) reduces to the equality in (\ref{disuguaglianza visibilità predicabilità}).

The CCR have been generalized by Jakob and Bergou for generic pure di-partite system  of arbitrary dimensions \cite{R_CCR_qudit} and recent studies by Basso and Maziero tried to generalize the relations for multipartite pure systems \cite{R11} and for mixed di-partite systems \cite{R12}. In these cases they used an alternative form of CCR  which includes the von Neuman entropy as measurement of the entanglement. For a pure state described by $\rho$, the CCR for the subsystem $k$ is \cite{R11}
\begin{equation}
\mathcal{C}_{re}(\rho^{k}) + \mathcal{P}_{vn}(\rho^{k}) + \mathcal{S}_{vn}(\rho^{k}) = \log_{2}d_{k}
\end{equation}
where the $\mathcal{S}_{vn}(\rho^{k})$ is the von Neumann entropy and it is related to the entanglement between $k$ and the rest of the system, and 
\begin{equation}
\mathcal{C}_{re}(\rho^{A_{1}}) = Tr[\rho^{A_{1}} \log_{2}\rho^{A_{1}} - \rho^{A_{1}} \log_{2}\rho^{A_{1}}_{diag}]
\end{equation}
is the relative entropy of coherence and it is related to the visibility, with $\rho^{k}_{diag} = \sum_{i_{k}} \rho_{i_{k} i_{k}}^{k} \ket{i_{k}}\bra{i_{k}}$ and 
\begin{equation}
\mathcal{P}_{vn}(\rho^{k}) = \log_{2}d_{k} - \mathcal{S}_{vn}(\rho^{k})
\end{equation}
is a measure of predictability.

In the case of a mixed system described by $\rho$, the CCR can be generalized by \cite{R12}
\begin{equation}
\mathcal{S}_{k|B}(\rho)+\mathcal{P}_{vn}(\rho^{k})+\mathcal{C}_{re}(\rho^{k})+\mathcal{I}_{k:B}(\rho) = \log_{2}d_{k}
\label{general_CCR_mixed}
\end{equation}
where $\mathcal{I}_{k:B}(\rho)=\mathcal{S}_{vn}(\rho^{k}) + \mathcal{S}_{vn}(\rho^{B}) - \mathcal{S}_{vn}(\rho)$ is called mutual information which is related to the entanglement between $k$ and $B=1 \cup 2 \cup k-1 \cup k+1 \cup... \cup n$ and $ \mathcal{S}_{k|B}(\rho) = \mathcal{S}_{vn}(\rho)-\mathcal{S}_{vn}(\rho^{B})$ is such that $\mathcal{S}_{k>B}(\rho) = - \mathcal{S}_{k|B}(\rho)$ is a measure of what we know less about a part of the system than we do about its whole.
They also studied the Lorentz invariance properties of CCR \cite{R13}.

\section{Formulas for the coefficients appearing in the lepton-antineutrino state}

The coefficients appearing in the lepton-antineutrino state Eq.~(\ref{eq:fullstate}) are given by
\begin{equation}
\begin{array}{rl}
\gamma_{1}(c,c') = & A (\omega_{1}^{(1)} (c,c') \cos^{2} \theta + \omega_{2}^{(1)}(c,c') \sin^{2}\theta ), \\
\gamma_{2}(c,c') = & A \omega^{(2)}(c,c') \sin\theta \cos\theta, \\
\gamma_{3}(c,c') = & B (\omega_{1}^{(3)} (c,c') \cos^{2} \theta + \omega_{2}^{(3)}(c,c') \sin^{2}\theta ), \\
\gamma_{4}(c,c') = & B \omega^{(4)}(c,c') \sin\theta \cos\theta
\end{array}
\end{equation}

where $A$ and $B$ are given by $(\ref{Coefficienti A e B})$ and
\begin{eqnarray}
\omega_{1}^{(1)} (c,c') &=& \bigg( \displaystyle\sum_{d} N_{\bar{\nu}_{1} }^{2} c^{\frac{d-1}{2}}f_{d}^{\bar{\nu}_{1}}f_{dc}^{\bar{\nu}_{1}} e^{-idE_{\bar{\nu}_{1}}t} \bigg) \bigg( \displaystyle\sum_{d'} d' N_{l}^{2} c'^{\frac{d'-1}{2}}f_{-d'}^{l}f_{d'c'}^{l} e^{-id'E_{l}t} \bigg), \nonumber \\
\omega_{2}^{(1)} (c,c') &=& \bigg( \displaystyle\sum_{d} N_{\bar{\nu}_{2} }^{2} c^{\frac{d-1}{2}}f_{d}^{\bar{\nu}_{2}}f_{dc}^{\bar{\nu}_{2}} e^{-idE_{\bar{\nu}_{2}}t} \bigg) \bigg( \displaystyle\sum_{d'} d' N_{l}^{2} c'^{\frac{d'-1}{2}}f_{-d'}^{l}f_{d'c'}^{l} e^{-id'E_{l}t} \bigg), \nonumber \\
\omega^{(2)} (c,c') &=& \bigg[ \displaystyle\sum_{d} \bigg( N_{\bar{\nu}_{1} }^{2} c^{\frac{d-1}{2}}f_{d}^{\bar{\nu}_{1}}f_{dc}^{\bar{\nu}_{1}} e^{-idE_{\bar{\nu}_{1}}t} - N_{\bar{\nu}_{2} }^{2} c^{\frac{d-1}{2}}f_{d}^{\bar{\nu}_{2}}f_{dc}^{\bar{\nu}_{2}} e^{-idE_{\bar{\nu}_{2}}t} \bigg) \bigg] \nonumber \\ 
&& \quad \bigg( \displaystyle\sum_{d'} d' N_{l}^{2} c'^{\frac{d'-1}{2}}f_{-d'}^{l}f_{d'c'}^{l} e^{-id'E_{l}t} \bigg).
\end{eqnarray}

and 
\begin{equation}
\begin{array}{l}
\omega_{1}^{(3)} (c,c') = -c c' (\omega_{1}^{1}(c,c'))^{*}, \\
\omega_{2}^{(3)} (c,c') = -c c' (\omega_{2}^{1}(c,c'))^{*}, \\
\omega^{(4)} (c,c') = -c c' (\omega^{2}(c,c'))^{*}
\end{array}
\end{equation}
with the notation $\bar{\nu}_{i} = \bar{\nu}_{m_{i}}$ for $i=1,2$.

\end{document}